\documentclass[prl,aps,twocolumn,superscriptaddress]{revtex4-1}
\usepackage{graphicx,color}
\usepackage{amsthm}
\usepackage{amsfonts}
\usepackage{algorithmic}
\usepackage{enumerate}
\usepackage{latexsym}
\usepackage{amsmath}
\usepackage{amssymb}
\usepackage{multirow}
\usepackage{array}
\usepackage{diagbox}
\usepackage{subfigure}
\usepackage[colorlinks=true,citecolor=blue,linkcolor=blue]{hyperref}

\def\avg#1{\left\langle#1\right\rangle}

\emergencystretch=\maxdimen
\begin{document}
\title{Evolution of magnetic correlation in an inhomogeneous square lattice}
\author{Xiao Zhang}
\affiliation{Department of Physics, Beijing Normal University, Beijing
100875, China}
\author{Runyu Ma}
\affiliation{Department of Physics, Beijing Normal University, Beijing
100875, China}
\author{Zenghui Fan}
\affiliation{Department of Physics, Beijing Normal University, Beijing
100875, China}
\author{Zixuan Jia}
\affiliation{Department of Physics, Beijing Normal University, Beijing
100875, China}
\author{Lufeng Zhang}
\email{lfzhang@bupt.edu.cn}
\affiliation{School of Science, Beijing University of Posts and Telecommunications, Beijing 100876, China}
\author{Tianxing Ma}
\email{txma@bnu.edu.cn}
\affiliation{Department of Physics, Beijing Normal University, Beijing
100875, China}
\affiliation{Key Laboratory of Multiscale Spin Physics(Ministry of Education), Beijing Normal University, Beijing 100875, China\\}

\begin{abstract}
We explore the magnetic properties of a two-dimensional Hubbard model on an inhomogeneous square lattice, which provides a platform for tuning the bandwidth of the flat band. In its limit, this inhomogeneous square lattice turns into a Lieb lattice, and it exhibits abundant properties due to the flat band structure at the Fermi level.
By using the determinant quantum Monte Carlo simulation, we calculate the spin susceptibility, double occupancy, magnetization, spin structure factor, and effective pairing interaction of the system.
It is found that the antiferromagnetic correlation is suppressed by the inhomogeneous strength and that the ferromagnetic correlation is enhanced.
Both the antiferromagnetic correlation and ferromagnetic correlation are enhanced as the interaction increases.
It is also found that the effective $d$-wave pairing interaction is suppressed by the increasing inhomogeneity.
In addition, we also study the thermodynamic properties of the inhomogeneous square lattice, and the calculation of specific heat provide good support for our point.
Our intensive numerical results provide a rich magnetic phase
diagram over both the inhomogeneity and interaction.
\end{abstract}
\maketitle

\section{Introduction}
Originally, the Lieb lattice was believed to be a paradigmatic model that could be used to characterize flat-band systems\cite{PhysRevLett.62.1201}, and it was explored as a route to itinerant ferromagnetism and superconducting and topological properties\cite{PhysRevLett.62.1201,PhysRevA.80.063622,PhysRevA.83.063601,PhysRevB.90.094506,Tsai_2015,PhysRevB.92.235106}.
In particular, tight-binding models with nearest-neighbor interactions based on the Lieb lattice have been discussed in the context of CuO$_2$ planes, especially in doped cuprates\cite{PhysRevLett.58.2794,doi:10.1126/science.235.4793.1196, doi:10.1126/science.288.5465.468,PhysRevB.55.14554,2017Experimental}.
The Lieb lattice presents a more accurate
three-band picture, which includes not only the square lattice of copper $d$ orbitals but also the intervening oxygen $p$
orbitals. For the copper dioxide plane (Cu-O) correlated with heavy metal atoms, the Cu-O plane is highly correlated with copper-based superconductors, and the heavy metal atoms above and below the copper dioxide plane are considered to be a reservoir for adjusting the density of electron and hole particles\cite{PhysRevB.90.140504}.
The Cu-O plane weakly coupled with these heavy metal atoms, and the coupling strength can be adjusted since the component and type of heavy metal atoms are experimentally modulated. Moreover, it is found that at the $M$ point of the Brillouin zone in twisted bilayer graphene systems the flat band intersects with a Dirac cone, which makes the Lieb lattice a possible model that can be used to characterize the superconducting behavior of twisted bilayer graphene\cite{PhysRevLett.123.237002,PhysRevB.101.060505,PhysRevLett.124.167002,PhysRevLett.125.030504}. In addition to the theoretical study of the Lieb lattice, flat band lattices have also been experimentally realized in optical lattices\cite{PhysRevB.81.041410,Guzm_n_Silva_2014,Zong_16,PhysRevLett.116.183902,PhysRevA.90.043624}. The optical lattice system allows the formation of a Lieb lattice with ultracold atoms, not only fermions\cite{PhysRevB.81.041410,Taie_2020} but also bosons\cite{doi:10.1126/sciadv.1500854,PhysRevLett.118.175301}.
Recently, tunneling coupled optical tweezer arrays in optical lattices\cite{PhysRevLett.128.223202,PhysRevLett.129.123201} might provide another promising platform to tune the flat band like twisted bilayer graphene and Cu-O. Current experimental technology allows us to excite atoms to a higher energy level, which makes it possible to realize the theoretical hypothesis.

\begin{figure}[htpb]
\centerline {\includegraphics*[width=4.5in]{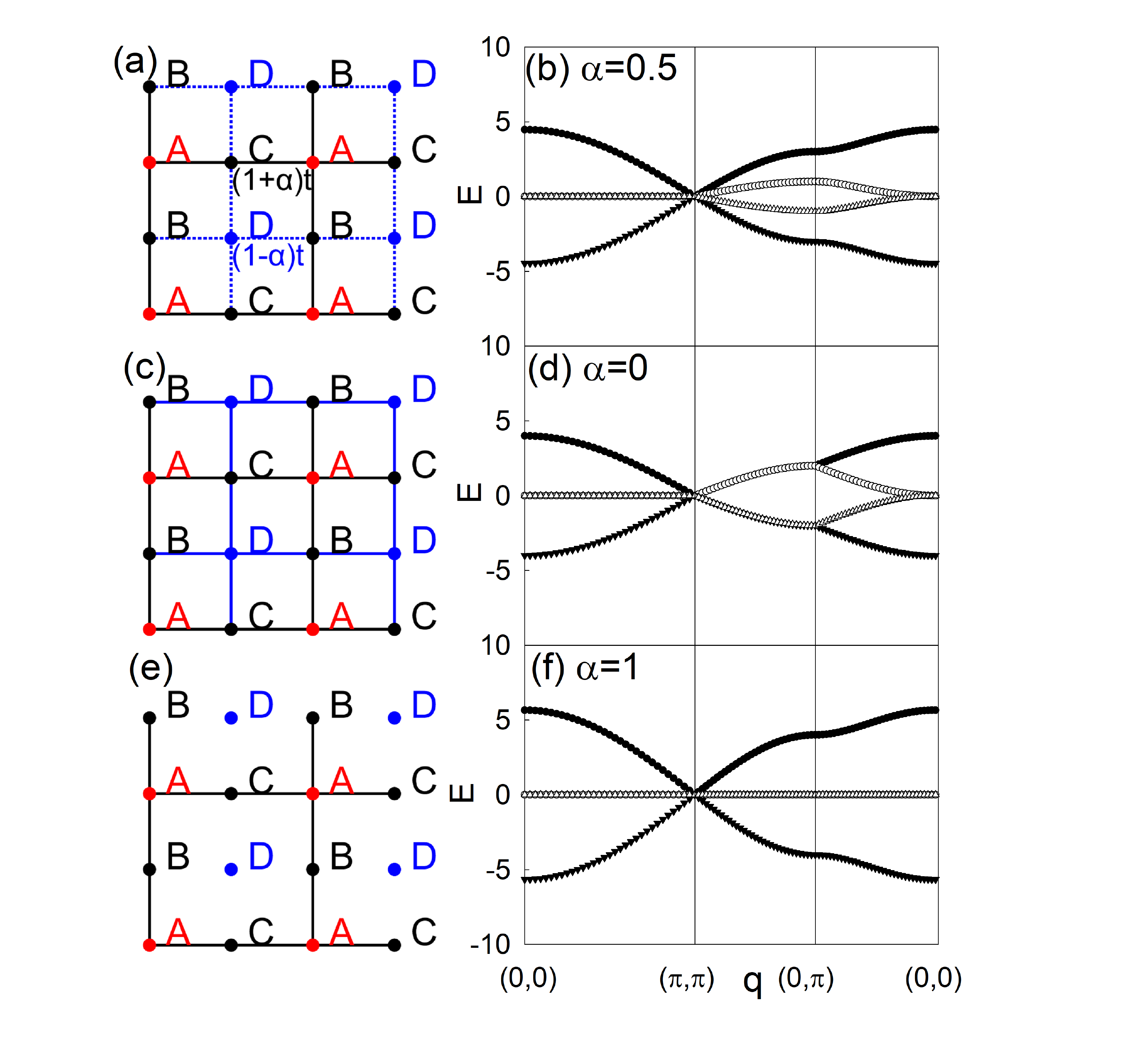}}
\caption{(Color online)
(a) A lattice with $L=4$ and the lattice inhomogeneity is introduced by modulated hopping. The smallest unit cell contains four types of sites, labeled A, B, C, and D. A sites have solid line bonds and D sites have dash line bonds, and B/C sites have both bonds.  The solid black lines bonds represent the hopping amplitude $(1+\alpha)t$, while the dashed blue line bonds represent $(1-\alpha)t$. When $\alpha=0$, the lattice become (c)square lattice, and the blue and black line both are $t=1$. When $\alpha=1$, the lattice become (e)Lieb lattice, and the blue lines dash which mean $t=0$. The energy band along the high symmetry line in the unfolded Brillouin zone for different inhomogeneities: (b) $\alpha=0.5$, (d) $\alpha=0.0$, and (f) $\alpha=1.0$. }
\label{Fig1}
\end{figure}

Based on these studies, one interesting model, the Hubbard model, of an inhomogeneous square lattice has attracted intensive attention\cite{PhysRevB.100.125141}.
As shown in Fig. \ref{Fig1} (a), inhomogeneity is introduced via modulated lattice hopping. The solid lines represent the hopping amplitude $(1 + \alpha)t$, while the dashed lines represent $(1 - \alpha)t$. At $\alpha=0$, as shown in Fig. \ref{Fig1} (c), this lattice reduces to a square lattice. In its limit with $\alpha=1$, as shown in Fig. \ref{Fig1} (e), this inhomogeneous square lattice turns into a Lieb lattice, and it exhibits abundant properties due to the flat band structure at the Fermi level.
Thus, the inhomogeneity could be modulated by adjusting $\alpha$, which provides an opportunity to study how flat bands should enhance correlated physics, such as magnetic order and superconductivity.
This tunability can also be compared with that of twisted multilayer graphene, for which the width of the low-energy bands can be tuned by changing the twist angle.

The magnetic phase diagram for the square lattice has been studied intensively, and some consensus about this model has been reached.
For example, the first-order metal-insulator Mott transition in the half-filled paramagnetic state and an infinitesimal critical
coupling strength for the antiferromagnetic phase at
half-filling due to the nested Fermi surface has been   established\cite{PhysRevB.104.094524,Berret1998,doi:10.1146/annurev-conmatphys-090921-033948,doi:10.1146/annurev-conmatphys-031620-102024,Bohrdt_2021}.
For the Lieb lattice, a series of rigorous results have been achieved.
Lieb established a theorem stating that in a class of bipartite geometries in any spatial dimension
the ground state is ferromagnetic at half-filling, as long as the number of atoms of each sublattice is different.
Applied to the case of the Lieb lattice, its ground state should be identified with {\it ferrimagnetism}, which means that
although each sublattice is indeed ferromagnetic, there is antiferromagnetic ordering between every pair of nearest neighbors\cite{PhysRevLett.62.1201}.
There is an open question that needs further study: how does the magnetic order evolve from the square lattice to the Lieb lattice with increasing inhomogeneity?
Within the framework of dynamical mean-field theory, magnetization and $d$-wave superconductivity have been studied in an inhomogeneous square lattice, in which a crossover from Fermi-liquid to non-Fermi-liquid behavior from dispersive to flat bands has been proposed\cite{PhysRevB.100.125141}.

In this paper, we explore the evolution of magnetic correlations in this inhomogeneous square lattice by using the determinant quantum Monte Carlo (DQMC) simulation. We are especially interested in the inhomogeneity-dependent ferromagnetic and antiferromagnetic correlation at half-filling. We calculate the thermodynamic specific heat, which helps us to further understand the evolution of magnetic correlations.
We also identify the dominant superconducting pairing symmetry in such an inhomogeneous square lattice.

\section{Model and method}
The Hamiltonian we studied is the Hubbard model on an inhomogeneous square lattice,
\begin{equation}
\begin{aligned}
H=&-\sum_{\langle ij \rangle,\sigma}\Big[(t_{ij} c_{i,\sigma}^\dagger c^{\phantom{\dagger}}_{j,\sigma}+ \text{h.c.}) \Big] -\mu N \\
&+U\sum\limits_{j} \left(n_{j,\uparrow}-\frac{1}{2}\right)\left(n_{j,\downarrow}-\frac{1}{2}\right),
\label{eq1}
\end{aligned}
\end{equation}
where $c_{i \sigma}^\dagger(c_{i \sigma})$ is the operator that creates (annihilates) an electron with spin $\sigma$ at site $i$.
To describe the inhomogeneity between the square and Lieb lattice,
we modulated the next-nearest hopping by setting $t_x=t[1+(-1)^y\alpha]$ and $t_y=t[1+(-1)^x\alpha]$, as shown in Fig. \ref{Fig1}(a),
which is the same as, the solid lines represent
the hopping amplitude $(1 + \alpha)t$  and the dashed lines represent $(1 - \alpha)t$.
In the second term, $\mu$ is the chemical potential, and the total particle number $N$ is defined as $N=\sum\limits_{ i,\sigma } c_{i,\sigma}^\dagger c^{\phantom{\dagger}}_{i,\sigma}$.
The last term in the Hamitonian introduces the on-site Hubbard interaction, where $U\ge0$ is the on-site interaction strength.
As inhomogeneity is introduced in the model by $\alpha$
, the average hopping strength is being kept.
The smallest unit cell contains four types of sites,
labeled A, B, C, and D, represented in Fig. \ref{Fig1}.
due to the flat band structure at the Fermi level.

At half-filling, particle-hole symmetry holds even in the inhomogeneous case, and the properties of the Hamiltonian Eq. \ref{eq1} could be solved using the DQMC method free of the infamous ``minus-sign problem".
Away from half-filling, DQMC simulations are plagued by
the sign problem, preventing us from
reaching very low temperatures. Nonetheless, we can still shed
some light on the effects on superconductivity.
In the present simulations, 4000$\sim$8000 sweeps were used to equilibrate the system,
and an additional 30000$\sim$80000 sweeps were made, each of which generated a measurement.

To study the magnetic correlation, we computed the spin structure factor, which is defined as\cite{PhysRevB.95.075142}
\begin{equation}
S(q)=\frac{1}{{N_s}}\sum_{i,j}e^{iq\cdot(i-j)}\left\langle S_i\cdot S_j\right\rangle,
\end{equation}
We compute the spin susceptibility in the $z$ direction at zero frequency, which is defined as \cite{doi:10.1063/1.3485059}
\begin{equation}
\chi(q)=\frac{1}{{N_s}}\int_0^\beta d \tau \sum_{i, j} e^{i q \cdot\left(i-j\right)}\left\langle m_i(\tau) \cdot m_j(0)\right\rangle,
\end{equation}
where $m_i(\tau)=e^{H \tau} m_i(0) e^{-H \tau}$ with $m_i=c_{i \uparrow}^{\dagger} c_{i \uparrow}-c_{i \downarrow}^{\dagger} c_{i \downarrow}$. $\chi$ is measured in units of $|t|^{-1}$, and $\chi(\Gamma)$ with $\Gamma=(0,0)$ measures the ferromagnetic correlation, while $\chi(K)$ with $K=(\pi,\pi)$ measures the antiferromagnetic correlation.
We also compute the double occupancy:
\begin{equation}
D_X=4\sum\limits_{j,X}(n_{j,\uparrow} n_{j,\downarrow}),
\end{equation}
and the local magnetization:
\begin{equation}
m_X=4\sum\limits_{j,X}(n_{j,\uparrow} + n_{j,\downarrow} -2n_{j,\uparrow} n_{j,\downarrow} ),
\end{equation}
where $X$ is the type of sites labeled A, B, C, and D.
We define the average magnetization for the cluster as $\bar{m} =(m_A+m_B+m_C+m_D)/4$
and the staggered magnetization is defined as $m_s=(m_A-m_B-m_C+m_D)/2$.
In this article, we discuss the situation at half filling by default, $\avg{n}=\avg{n_{\uparrow}}+\avg{n_{\downarrow}}=1.0$.

In general, the magnetic correlation can be well reflected by the above physical quantities. In the perspective of finite temperature thermodynamics, we can also find corresponding evidence. Especially in specific heat, the magnetic correlation is closely related to the specific heat peak, which is adequately understood in square lattice\cite{PhysRevB.63.125116} and honeycomb lattice\cite{PhysRevB.72.085123}. To further understand the evolution of magnetic correlations in the inhomogeneous square, we also perform calculations for the thermodynamic specific heat. The specific heat is calculated by differentiating a nonlinear fit of the energy as the form:
\begin{equation}
c(T)=\frac{\partial e_{fit}(T)}{{\partial T}},
\end{equation}
We start by calculating the energy date of the system by the DQMC method, and use an exponential fit of the energy
by the function:
\begin{equation}
e_{fit}(T)=c_{0}+\sum_{n=1}^{M}c_{n}e^{-\beta n\Delta},
\end{equation}
We choose a cutoff at $M = 7$.

\section{Results and discussion}

We first present the band structure of the inhomogeneous square lattice in its noninteracting limit.
In Fig. \ref{Fig1}(b)(d)(f), the energy bands along the high symmetry line in the unfolded Brillouin zone with different $\alpha$ are presented.
As the inhomogeneity factor $\alpha$ increases from $0.0$ to $0.5$ to $1.0$, the energy band is split from the $2$ to $4$ bands, then degenerated to $3$ bands to form a flat band, as shown in Fig. \ref{Fig1}(f).

\begin{figure}[htpb]
\centerline {\includegraphics[width=3.2in]{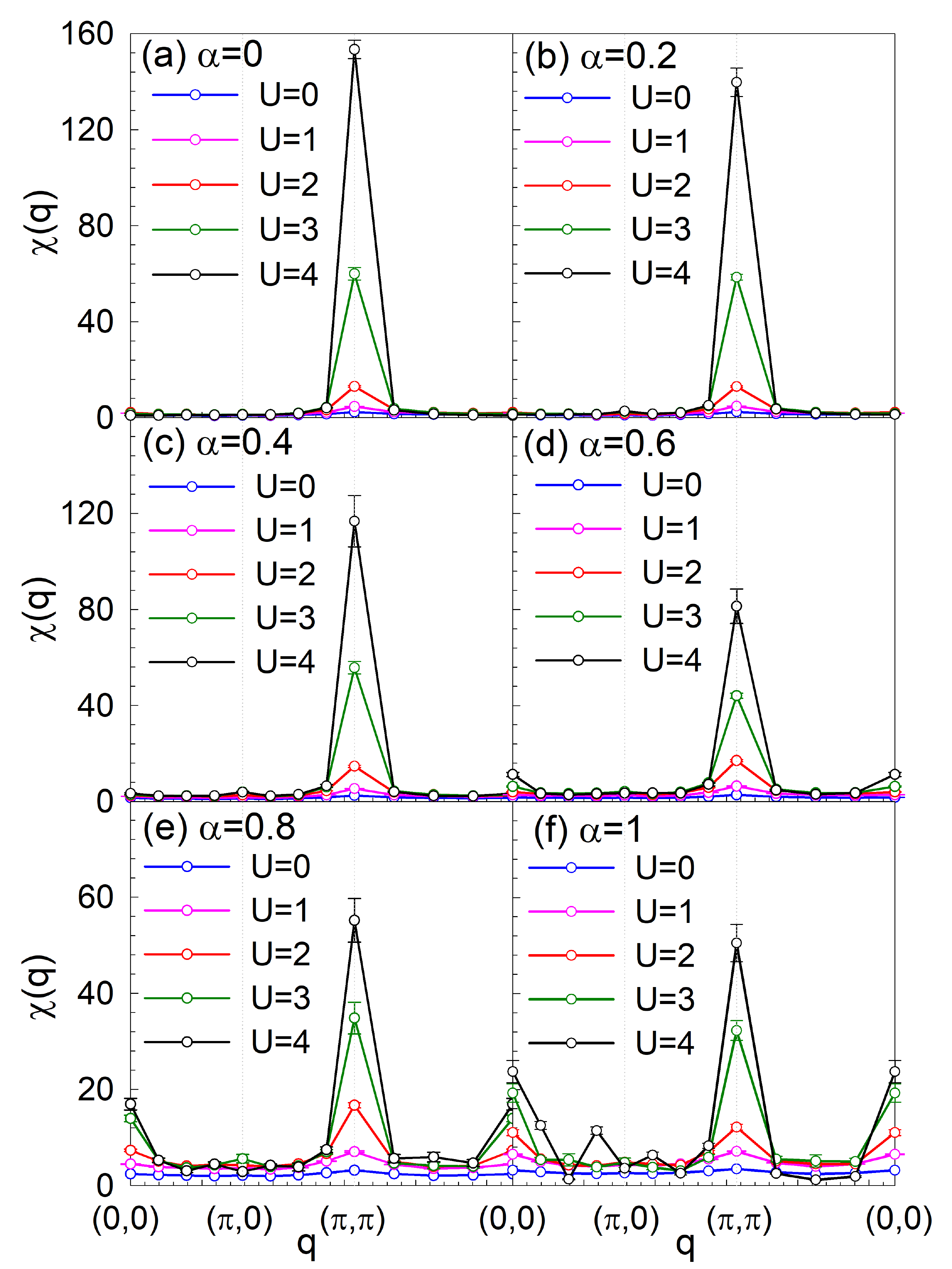}}
\caption{(Color online) Spin susceptibility $\chi(q)$ in momentum space along the high symmetry points in the Brillouin zone. Results are obtained at $\beta=6$ and $L=8$ and presented for different values of the inhomogeneity $\alpha$ and the interaction $U$.}
\label{Fig2}
\end{figure}

To explore how the magnetic properties change with the band structure,
we present the spin susceptibility $\chi(q)$ as a function of momentum $q$ with different $U$ at half-filling for various inhomogeneities $\alpha$ in Fig. \ref{Fig2}.
As Fig. \ref{Fig2} shown, for any inhomogeneity $\alpha$, the spin susceptibility increases as the interaction $U$ increases, especially at the peak $q=K=(\pi,\pi)$, indicating that the antiferromagnetic correlation is robust in the system, either for a square lattice or Lieb lattice. As $\alpha$ increases, the correlation at the $\Gamma=(0,0)$ point arises, which corresponds to the ferromagnetism of the Lieb lattice at half-filling.
When $\alpha=0.2,0.4$, the peak value of $U=3$ at the $K$ point is reduced compared to $\alpha=0.0$, while the values corresponding to the other vectors have a certain increment.
When $\alpha=0.6$, there is also a peak at the $\Gamma$ point, but it is still smaller than that at $\chi(K)$, and as the value of $\alpha$ increases to $1$, the difference between the two peaks decreases.
We also notice that $\chi(K)>\chi(\Gamma)$ when $\alpha=1$.
\begin{figure}[htpb]
\centerline {\includegraphics*[width=3.2in]{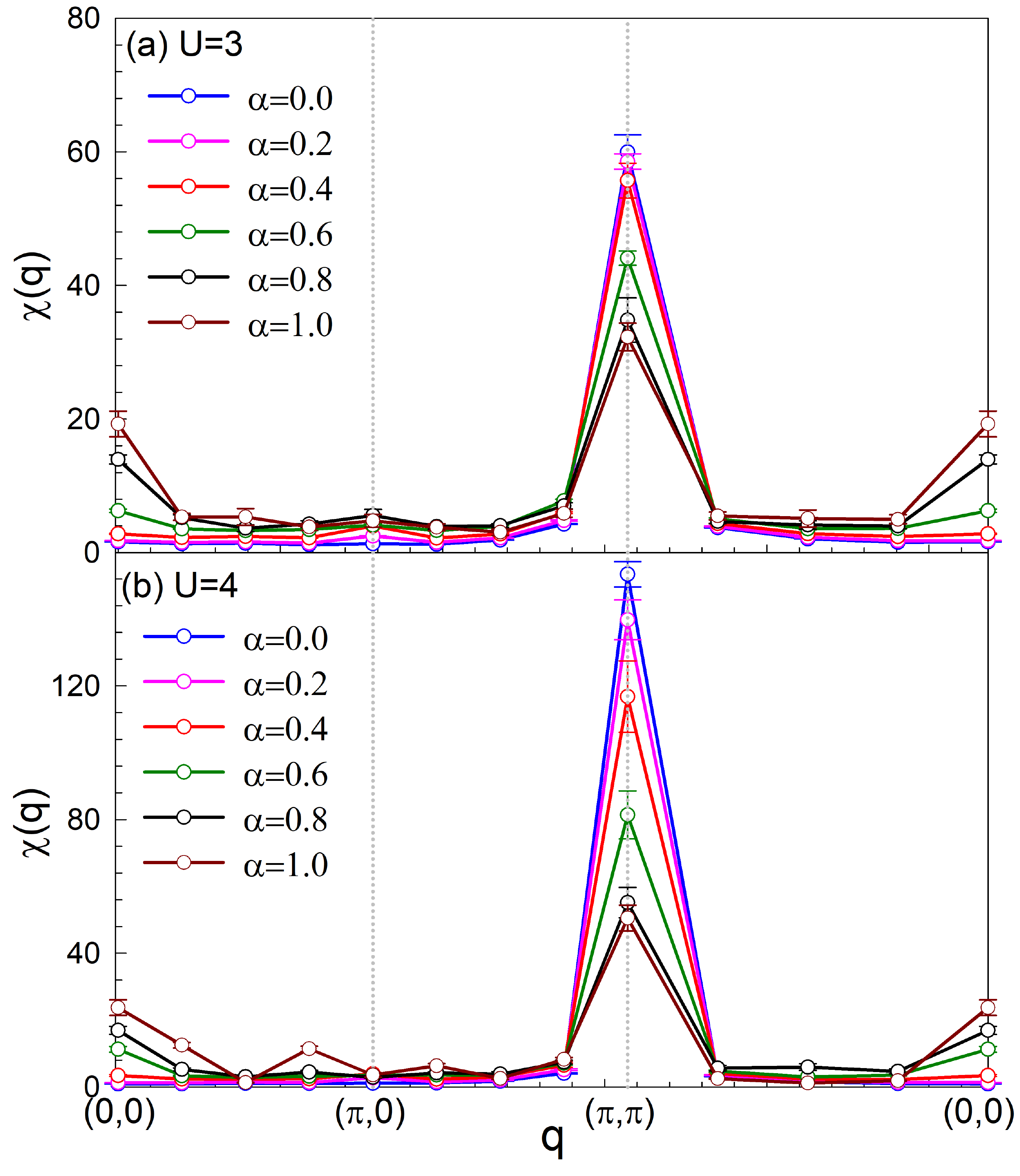}}
\caption{(Color online) Spin susceptibility $\chi(q)$ in momentum space along the high symmetry points in the Brilloiun zone. Results are obtained at $\beta=6$ and $L=8$ and presented for different values of the inhomogeneity $\alpha$ for (a) $U/t=3$ and (b) $U/t=4$.}
\label{Fig3}
\end{figure}

The effect of $\alpha$ is more apparent when we fix the interaction strength $U$. As Fig. \ref{Fig3} shows, $\chi(q)$ is plotted as a function of momentum $q$ with different $\alpha$ at interaction strengths $U=3.0$ (a) and $U=4.0$ (b) for a fixed temperature $\beta=6$. We can see that the peak at $ K$ decreases as $\alpha$ increases, while $\chi(\Gamma)$ slightly increases.

\begin{figure}[htpb]
\centerline {\includegraphics*[width=3.2in]{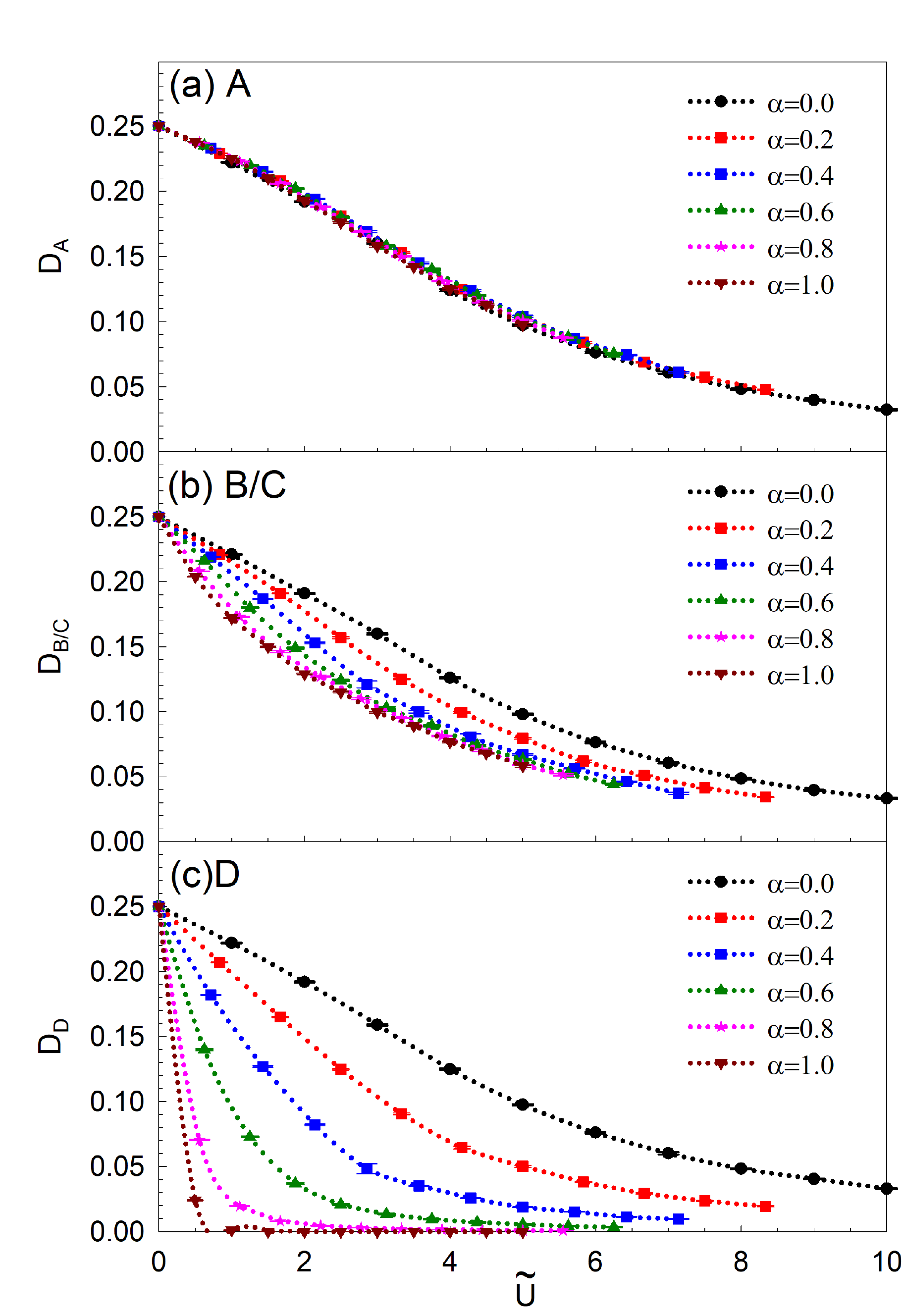}}
\caption{(Color online) (a) Double occupancy on $A$ sites for varying interaction $\tilde{U}$ with different $\alpha$ at $\beta=6$ and $L=8$, (b) double occupancy on $B/C$ sites and (c) double occupancy on $D$ sites.}
\label{Fig4}
\end{figure}

Then, we discussed the effect of the interaction and inhomogeneity on the double occupancy and magnetic moment at different kinds of lattice sites. In Fig. \ref{Fig4}, we compared the values of double occupancy corresponding to different homogeneity at inverse temperature $\beta=6$, and we set varying interaction values to $\tilde{U}=U/t(1+\alpha)$. The corresponding double occupancy value of lattice site A is shown in Fig. \ref{Fig4}(a), while values on sites B/C are shown in Fig. \ref{Fig4}(b).
In the homogeneous system, that is, when $\alpha=0.0$, the double occupancy curves behave the same at lattice sites A and B/C. However, when $\alpha$ increases, the double occupancy at the A site behaves the same way, as shown in Fig. \ref{Fig4}(a). On the other hand, the double occupancy on sites B/C has a significant decrease and behaves exponentially at small $\tilde{U}$. This may be caused by the fact that the flat band electrons only occupy the B/C sites, and the existence of the flat band supports single occupancy even for infinitesimal $\tilde{U}$.

As shown in Fig. \ref{Fig4} (c), compared with lattice sites A/B/C, the corresponding double occupancy value of D changes sharply with $\tilde{U}$ and $\alpha$, and when $\alpha=1.0$, the double occupancy declines to $0$ after $\tilde{U}=0.5$.
Briefly, as Fig. \ref{Fig4} shows, the double occupancy will tend to zero at all sites in the strong coupling limit regardless of $\alpha$.
Fig. \ref{Fig4} (c) also implies that we should pay attention to the influence of the magnetization of heavy metal atoms on the properties of copper-based superconductors.

\begin{figure}[htpb]
\centerline {\includegraphics*[width=3.2in]{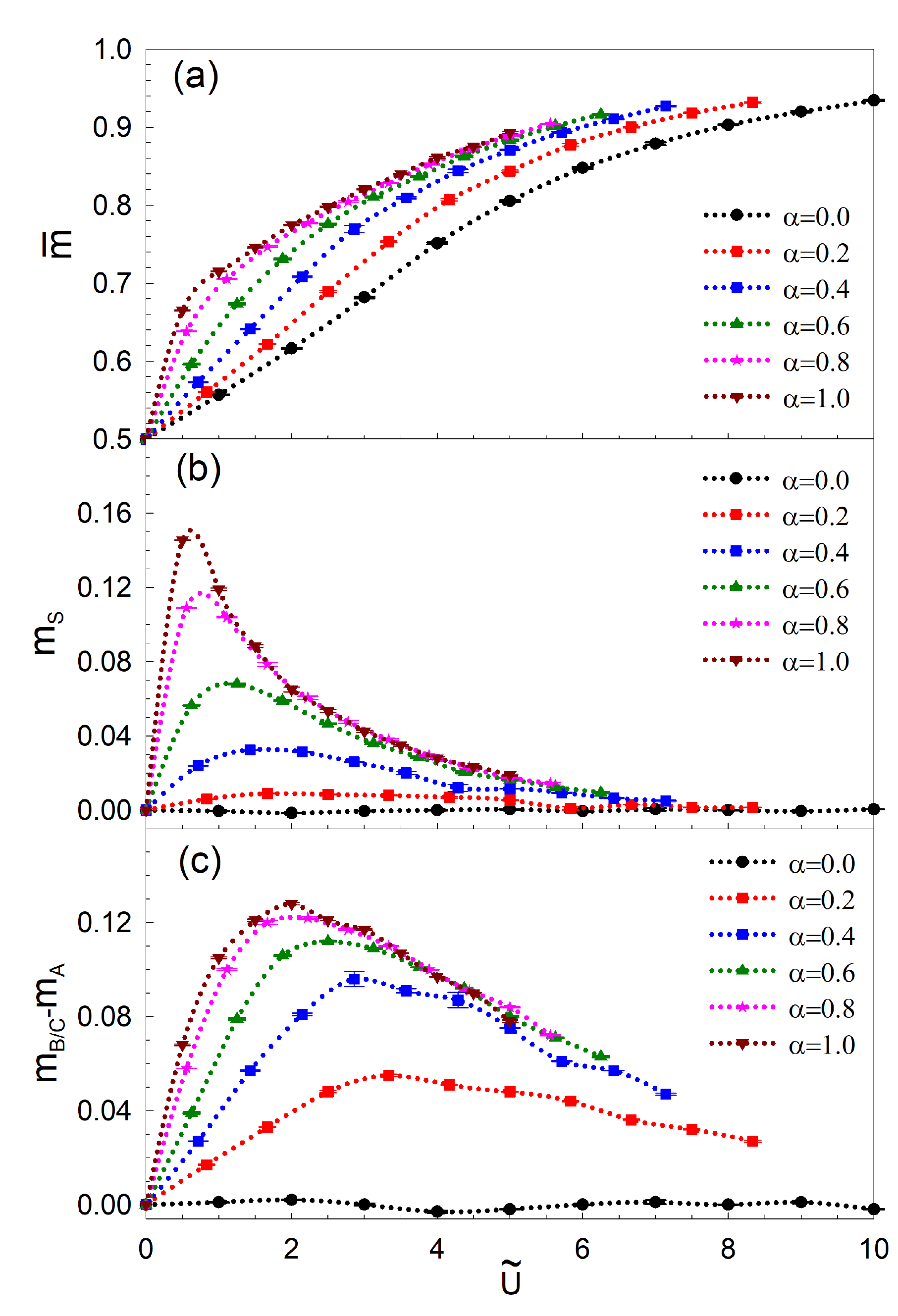}}
\caption{(Color online)
(a) Average magnetization of lattice $\overline{m}$, (b) Staggered magnetization of lattice $m_s$ and (c) magnetization $m_{B/C}-m_A$ for varying interaction $\tilde{U}$ for different $\alpha$ at $\beta=6$ and $L=8$. }
\label{Fig5}
\end{figure}

Then, we discuss the interaction of inhomogeneities on the uniform magnetization $m_{s}$ and average magnetization $\bar{m}$ in Fig. \ref{Fig5}.
As shown in
Fig. \ref{Fig5} (a), $\bar{m}$ increases as $\alpha$  and $U$ increases. While all the m values converged to the same limit for different $\alpha$, the uptrend of the curve is sharper with larger $U$ strength.
As shown in Fig. \ref{Fig5} (b), $m_{s}$ increases with increasing $U$ when $U<U_{p}$,
and then decreases, thus forming a peak at $U_p$. As $\alpha$
increases, we find that the peak moves toward lower $U$.
In the strongly interacting regime, all curves for different $\alpha$ coalesce and asymptotically approach zero.
In Fig. \ref{Fig5} (c), $m_{B/C}-m_{A}$ is considered as $m_{s}$ for the Lieb lattice. Compared with Fig. \ref{Fig5} (b), the behavior of the curves is roughly similar, except that the magnetization changes gently with $\tilde{U}$ but changes sharply with $\alpha$.
Our results on the double occupancy and uniform magnetization are in agreement with previous studies within dynamical mean-field theory\cite{PhysRevB.100.125141}.

\begin{figure}[htpb]
\centerline {\includegraphics*[width=3.2in]{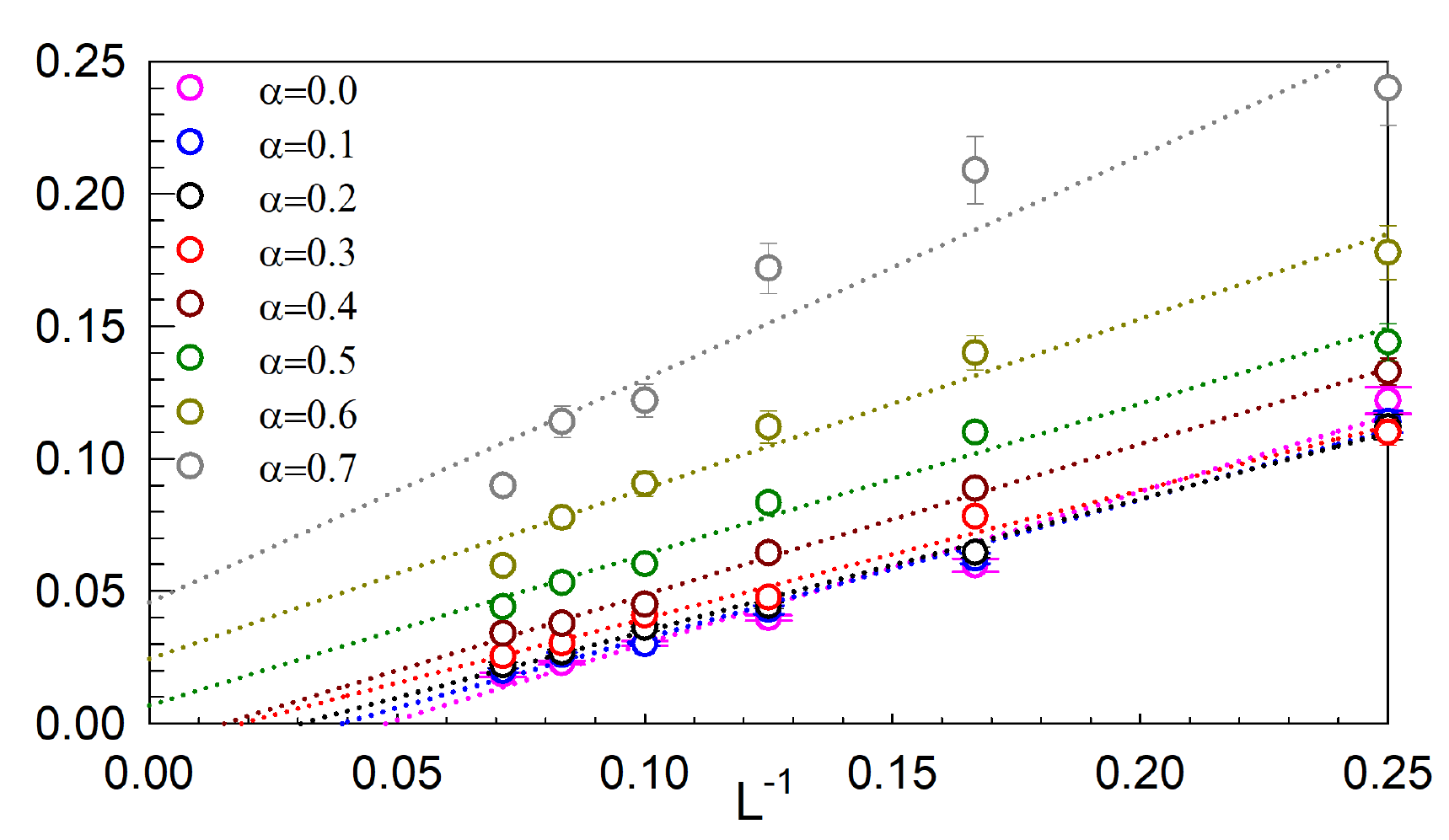}}
\caption{(Color online)
Finite-size scaling of the spin structure factor $S(\Gamma)$, which changes with the inhomogeneity $\alpha$ from 0 to 0.7, at interaction strengths $U=3$ and inverse temperature $\beta=6$. }
\label{Fig6}
\end{figure}

A more rigorous probe of long range order is accomplished using finite-size scaling analyses. The order parameter can be obtained by normalizing the structure factor $S(\Gamma)$ to the thermodynamic limit, as shown in Fig. \ref{Fig6}.
When inverse temperature $\beta=6$, according to linear fitting, it can be predicted that under the size limit, $S(\Gamma)$ will gradually increase with the increase of $\alpha$, and when it increases to $\alpha=0.5$, a positive value will appear, which means that there is a possible ferromagnetic order in the system when $\alpha$ larger than 0.5.

To elucidate the effect of inhomogeneity on the superconductivity, we also studied the effective pairing interaction as a function of inhomogeneity. Following previous studies\cite{PhysRevB.39.839,PhysRevB.40.506,PhysRevLett.110.107002}, pairing susceptibility is defined as
\begin{equation}
P_s = {\frac{1}{N_S}}{\sum_{ij}\int_{0}^{\beta}{\mathrm{d}\tau}\langle \Delta_{s}^\dagger(i,\tau)\Delta_{s}(j,0)\rangle},
\label{4}
\end{equation}
where $s$ is the pairing symmetry.
Due to the constraint of the on-site Hubbard interaction in Eq. (\ref{eq1}), the corresponding order parameter $\Delta_{s}^\dagger(i)$ is
\begin{equation}
\Delta_{s}^{\dagger}(i) = \sum_{\textbf{l}}f_{s}^{\dagger}(\delta_{\textbf{l}})(c_{i\uparrow}c_{i+\delta_{\textbf{l}}\downarrow}-c_{i\downarrow}c_{i+\delta_{\textbf{l}}\uparrow})^{\dagger},
\label{5}
\end{equation}
where $f_s(\delta_\textbf{l})$ is the form factor of the pairing function.
$P_s$ includes both the renormalization of the propagation of the individual particles and the interaction vertex between them, whereas $\widetilde{P}_s$ includes only the former effect. To extract the effective pairing interaction in a finite system, one should subtract from $P_s$ its uncorrelated single-particle contribution $\widetilde{P}_s$, which is achieved by replacing $\langle c_{i\downarrow}^{\dagger}c_{j\downarrow}c_{i+\delta_\textbf{l}\uparrow}^{\dagger}c_{j+\delta_{\textbf{l}^{'}}\uparrow}\rangle$ in Eq. (\ref{4}) with $\langle c_{i\downarrow}^{\dagger}c_{j\downarrow}\rangle \langle c_{i+\delta_\textbf{l}\uparrow}^{\dagger}c_{j+\delta_{\textbf{l}^{'}}\uparrow}\rangle$, and the effective pairing interaction $\textbf{P}_s$ is defined as $\textbf{P}_s=P_s-\widetilde{P}_s$. The positive $\textbf{P}_s$, namely, $\textbf{P}_s>0$, indicates the presence of superconductivity. More details can be found in Ref. \cite{PhysRevB.39.839,PhysRevB.40.506,PhysRevLett.110.107002}.

\begin{figure}[htpb]
    \centerline {\includegraphics*[width=3.2in]{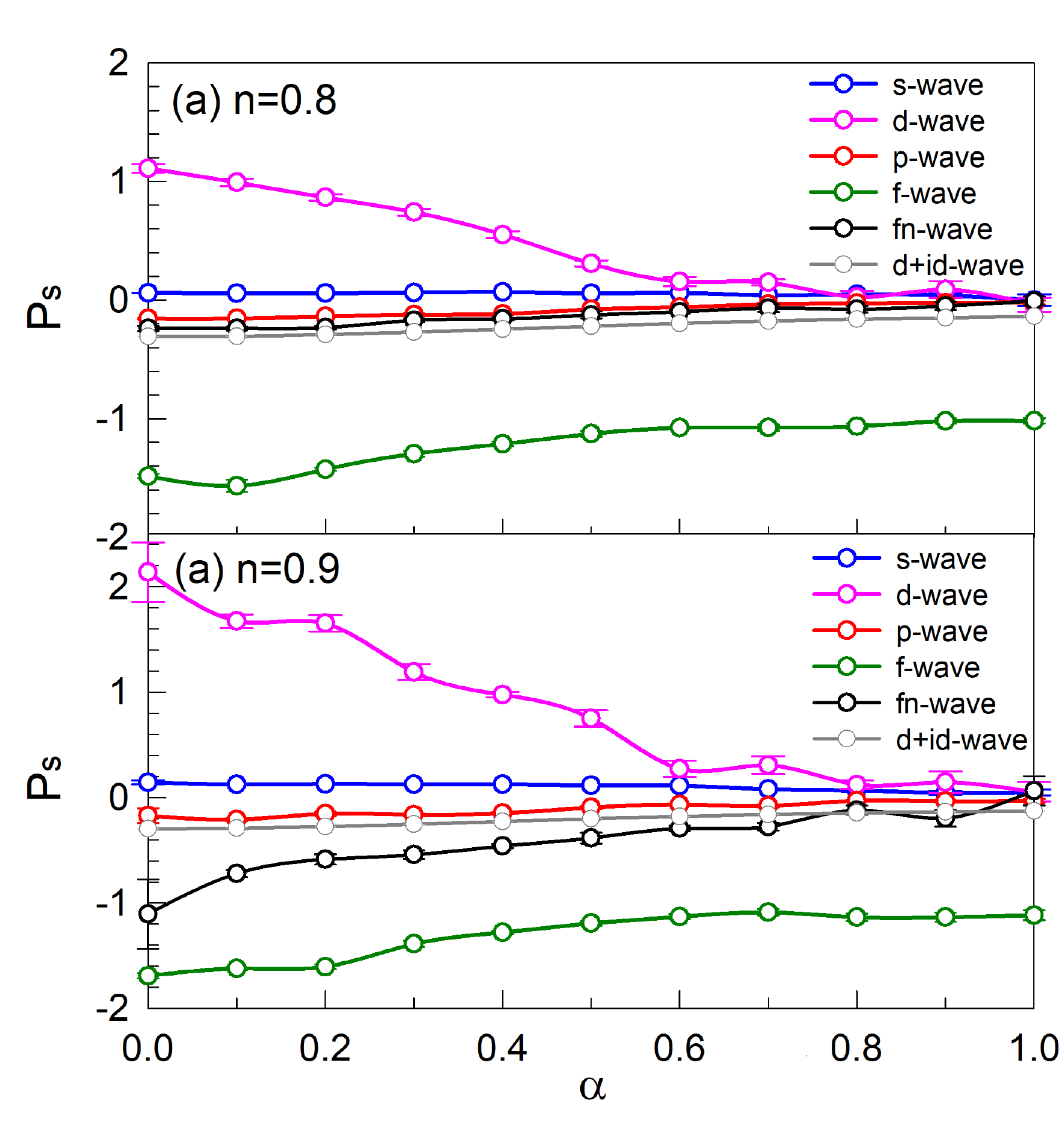}}
    \caption{(Color online) Effective pairing interaction $P_{s}$ as a function of inhomogeneity $\alpha$ for different pairing symmetries for (a) charge density $n$=0.8 and (b) charge density $n$=0.9 with $\beta=6$ and $L=8$.}
    \label{Fig:Pairing}
    \end{figure}

It is widely known that the dominant pairing symmetry is the $d$-wave in the Hubbard model on a square lattice\cite{doi:10.1063/1.4961462,PhysRevLett.80.5188}. As shown in Fig. \ref{Fig:Pairing}, the effective pairing interaction with different pairing symmetries is shown for (a) $n$=0.8 and (b) $n$=0.9 with $U=3.0$ and $\beta=6$. One can see that for both $n$=0.8 and $n$=0.9, the effective pairing interaction with the $d-$wave is positive, and the others are negative. This means that superconductivity with $d$-wave pairing symmetry is possible. Moreover, the effective pairing interaction decreases as the inhomogeneity $\alpha$ increases, and it tends to be zero as $\alpha=1.0$, which indicates that at least the $d$-wave superconductivity should be suppressed by the increasing inhomogeneity. The effective pairing interaction with the $d+id$ wave is not obvious with negative small value.

\begin{figure}[htpb]
\centerline {\includegraphics*[width=3.2in]{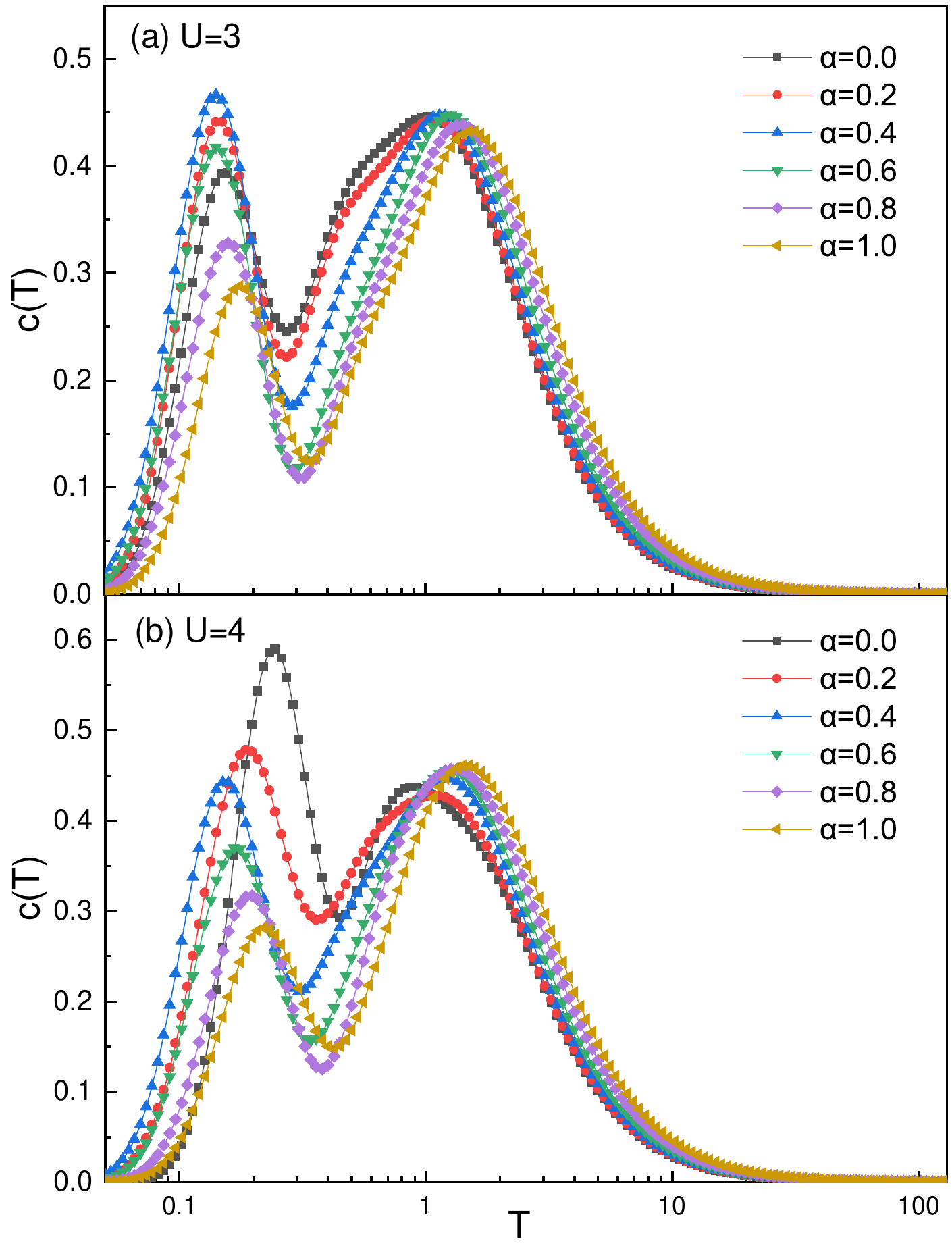}}
\caption{(Color online) Results for specific heat $c(T)$ calculated at $L=8$ by fitting method. Those results are presented for different values of the inhomogeneity $\alpha$ for (a) $U=3$ and (b) $U=4$.}
\label{Fig8}
\end{figure}

Finally, we use the same parameters as Fig.~\ref{Fig3}
to calculate the specific heat, which is shown in Fig.~\ref{Fig8}. The same parameters are chosen so that we can
directly compare the physical quantities of these two
figures, which provides useful support for our view
on the evolution of magnetic correlations with the
inhomogeneity $\alpha$.
Fig. \ref{Fig8} shows the specific heat as a function of temperature, inhomogeneity and interaction, with all cases showing a two-peak structure.
Regardless of $U=3$ or $U=4$, as $\alpha$ increases, the high-temperature peak associated with the generation of the local moment has a move to the high temperature region.
Here we put more attention on the low-temperature peak which is correlated with the collective spin excitation.
The increasing temperature tends to destroy the magnetic order,
and the position of low-temperature peak, $T_{low}$ indicating where a magnetic transition may develop.
The ground state magnetic correlation shall be more strong with a higher $T_{low}$.
As shown in Fig. \ref{Fig8} (a) for $U=3$, $T_{low}$ has no obvious change for $\alpha=0,0.2,0.4$ and starts to increase from $\alpha=0.6$,
suggesting that there is no obvious change in the antiferromagnetic correlation at $\alpha=0,0.2,0.4$ while ferromagnetic correlation is likely established and enhanced from $\alpha=0.6$.
Also in Fig. \ref{Fig8} (b) for $U=4$, with $\alpha=0,0.2,0.4$, $T_{low}$ has a significant decrease, corresponding to the suppression of antiferromagnetic correlation, and then increases like that of $U=3$.
The turning point approximately occurs at $\alpha=0.4\sim0.6$, indicating where the ferromagnetic correlation develops.
Compared with that of Fig.~\ref{Fig3}, we can find a good consistency between spin susceptibility $\chi(q)$ and specific heat $c(T)$.

In addition, for both $U=3$ and $U=4$, the low-temperature peak has a obvious move to the high temperature region with the increasing $\alpha$ as $\alpha>0.4$, while the peak value is decreasing.
It may suggest that ferromagnetic correlation tends to be strong.
Since the fiercer the competition between antiferromagnetic and ferromagnetic correlation in the ground state, the smaller the entropy change caused by spin excitation at low temperature,
resulting in a small low-temperature peak of specific heat.

\section{Conclusions}
\label{sec:conclusions}

In summary, by using the determinant quantum Monte Carlo method, we studied an inhomogeneous square lattice, which turns into a Lieb lattice in the inhomogeneous limit with $\alpha=1$. The special lattice provides us with a platform to study the features of flat-band structure systems. As $\alpha$ increases, the spin susceptibility $\chi(K)$ decreases, while $\chi(\Gamma)$ increases. In consideration of interactions, we find that $\chi(q)$ is enhanced when $U$ increases. Then, we studied the double occupancy, magnetization moment, relation between ferromagnetic order and $\alpha$, the effective pairing interaction  and the specific heat. Our intensive numerical results provide a global understanding of the evolution of magnetic correlations in an inhomogeneous square lattice.

\begin{acknowledgments}
This work is supported by NSFC (No. 11974049).
The numerical simulations were performed at the HSCC of Beijing Normal University and on the Tianhe-2JK supercomputer in the Beijing Computational Science Research Center.
\end{acknowledgments}

\setcounter{equation}{0}
\setcounter{figure}{0}
\renewcommand{\theequation}{A\arabic{equation}}
\renewcommand{\thefigure}{A\arabic{figure}}
\renewcommand{\thesubsection}{A\arabic{subsection}}

\section{Appendix}

\subsection{Visually Perceptive Colormap}
\label{appendix:A}
\begin{figure}[htpb]
  \centerline {\includegraphics*[width=3.2in]{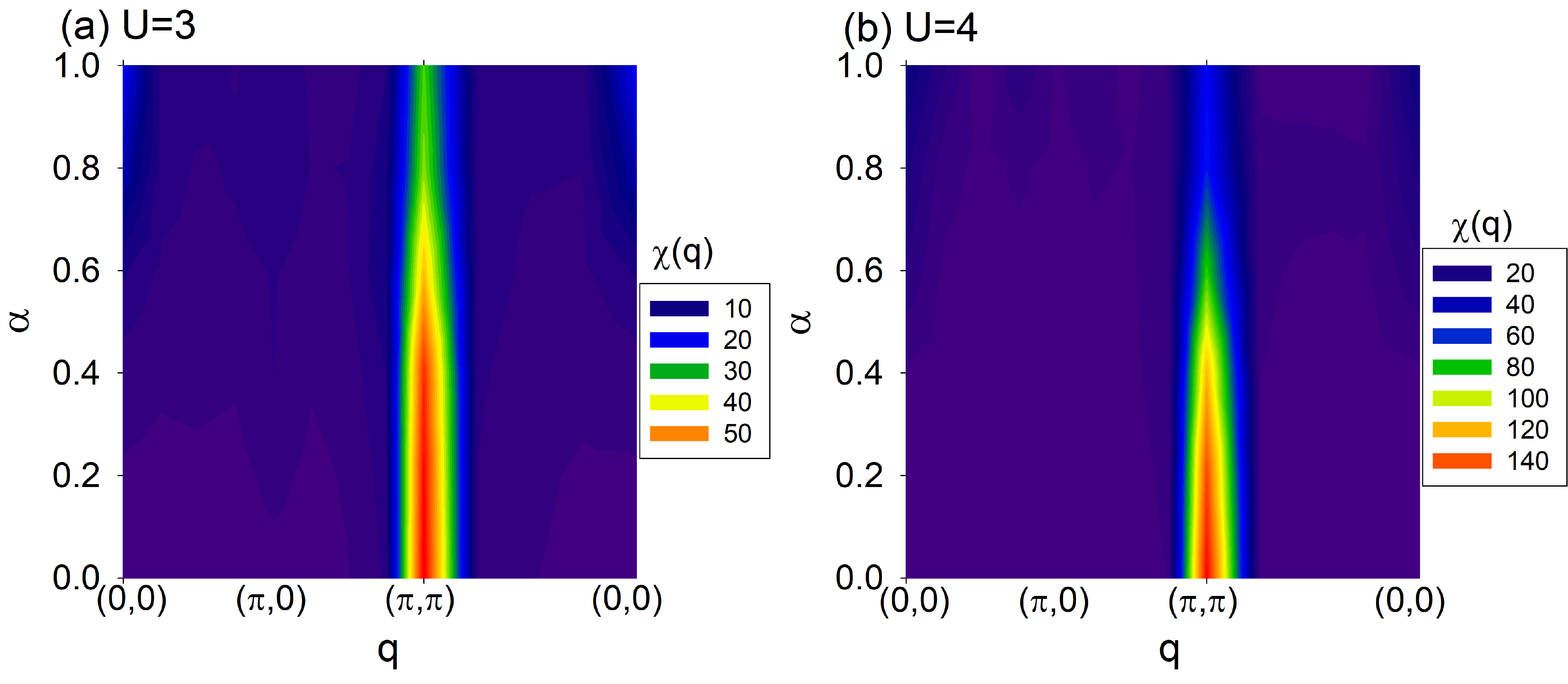}}
  \caption{(Color online) Colormap of spin susceptibility $\chi(q)$ in momentum space along the high symmetry lines in the Brilloiun zone. Results are presented for different values of the inhomogeneity $\alpha$ for (a) $U/t=3$ and (b) $U/t=4$ with $\beta=6$ and $L=8$.}
  \label{Fig3c}
  \end{figure}

\begin{figure}[htpb]
    \centerline {\includegraphics*[width=3.2in]{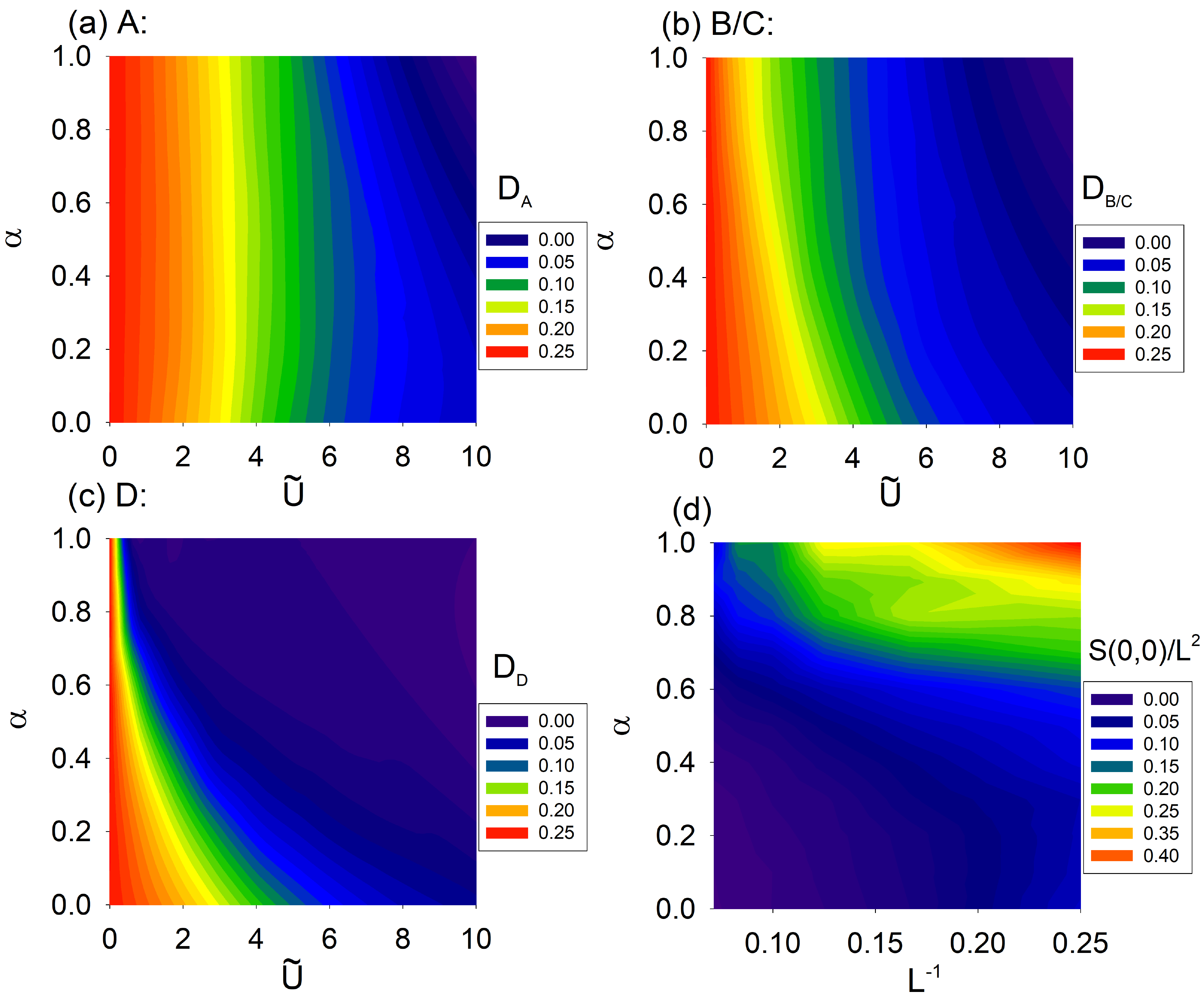}}
    \caption{(Color online) Colormap of(a) double occupancy on $A$ sites for varying interaction $\tilde{U}$ with different $\alpha$ at $\beta=6$ and $L=8$, (b) double occupancy on $B/C$ sites, (c) double occupancy on $D$ sites and (d) normalized spin structure factor $S(\Gamma)$ for lattice size scaling with different $\alpha$ at $U=3$ and $\beta=6$.}
    \label{Fig46c}
  \end{figure}

\begin{figure}[htpb]
  \centerline {\includegraphics*[width=3.2in]{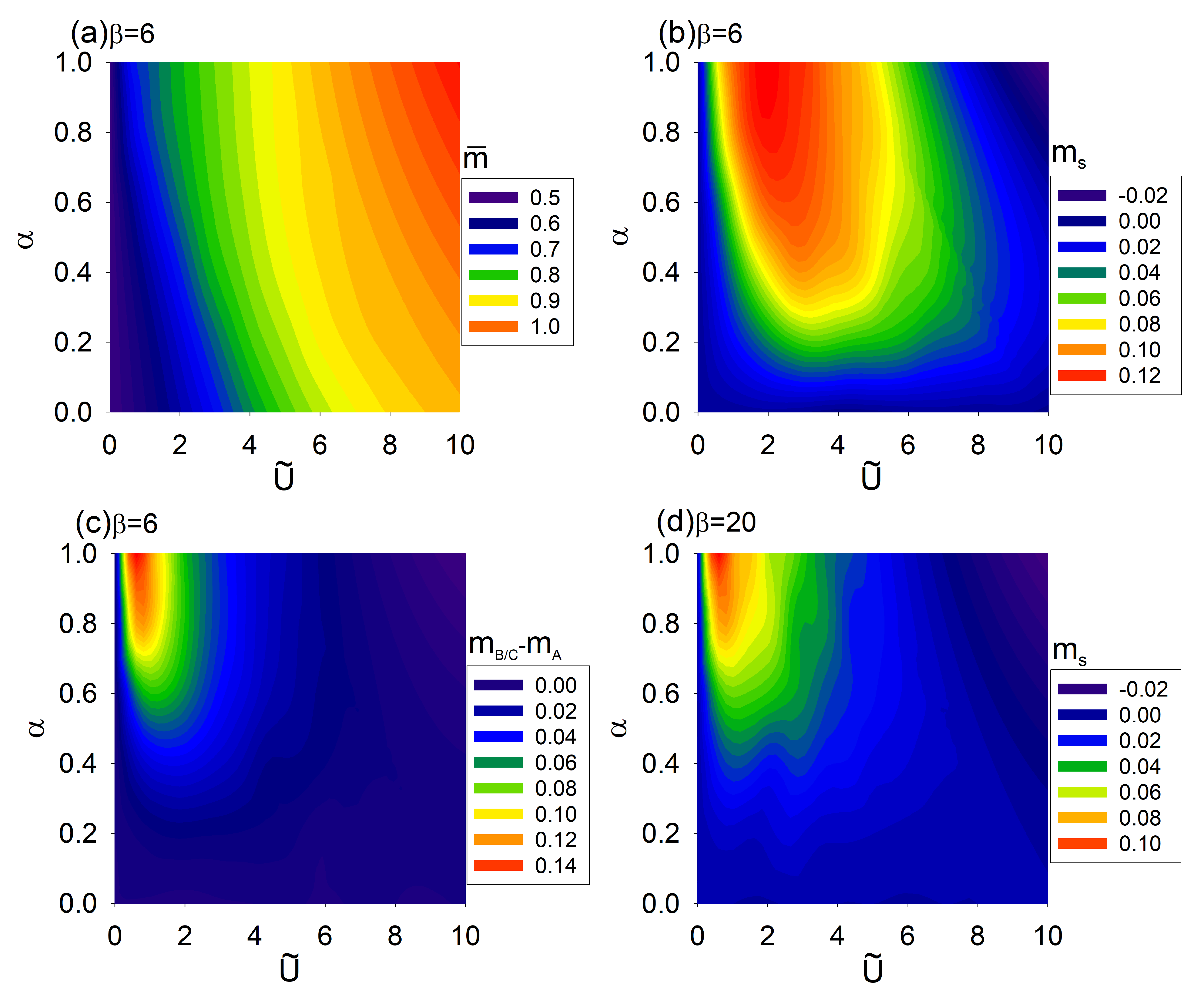}}
  \caption{(Color online)  Colormap of (a) average magnetization of lattice $\overline{m}$, (b) staggered magnetization of lattice $m_s$ and (c) magnetization $m_{B/C}-m_A$ for varying interaction $\tilde{U}$ with different $\alpha$ at $\beta=6$ and $L=8$. (d) is for staggered magnetization of lattice $m_s$ at $\beta=20$.}
  \label{Fig5c}
  \end{figure}

The use of a visually perceptive colormap is useful since it can carry informational content.
Visually perceptive colormaps are present for Figs. \ref{Fig3}-\ref{Fig6} , which are shown in Figs. \ref{Fig3c}, \ref{Fig46c}, \ref{Fig5c}, to have a better view of these order parameters correspond to smaller or larger value of $\alpha$. As shown in Fig. \ref{Fig3c}, we could see how spin susceptibility $\chi(q)$ change with $q$ for different values of $\alpha$. In Figs. \ref{Fig46c}(a)(b)(c), how double occupancy $D$ changes with $\tilde{U}$ for different values of $\alpha$ are shown. Fig. \ref{Fig46c}(d) shows how normalized spin structure factor $S(\Gamma)$ changes with $L^{-1}$ for different values of $\alpha$. In Fig. \ref{Fig5c}, we could see how magnetization $m$ change with $\tilde{U}$ for different values of $\alpha$.

\subsection{Numerical simulations for larger $U$ and $\beta$}
\label{appendix:B}
\begin{figure}[htpb]
\centerline {\includegraphics*[width=3.0in]{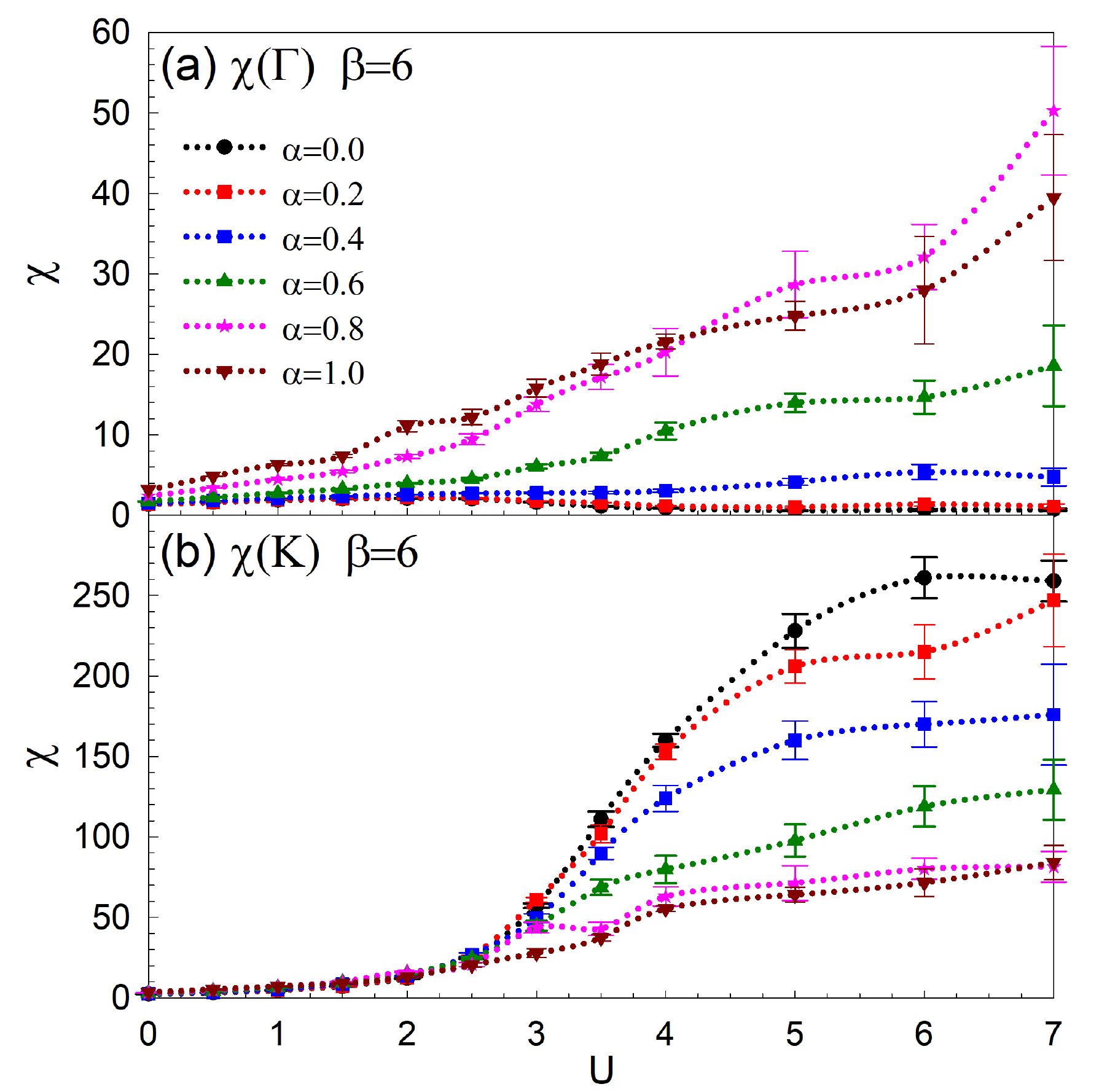}}
\caption{(Color online) Antiferromagnetic and ferromagnetic spin susceptibility $\chi(K)$, $\chi(\Gamma)$ presented for different values of the inhomogeneity $\alpha$ and the interaction $U$. Results are obtained at $\beta$=6 and $L=8$.}
\label{figm3}
 \end{figure}

In general, antiferromagnetism is enhanced around $U/t=8\sim12$ in two- and three-dimensional
square lattices\cite{doi:10.1146/annurev-conmatphys-090921-033948,PhysRevA.102.033340,PhysRevB.94.125114.}
It is questionable on the behavior of magnetic order in this inhomogeneous square lattice for larger interactions and lower temperatures.
We have extended numerical simulation for the spin susceptibility to some larger interaction strength, from which one can see that our main conclusion remains unchanged. In Fig. \ref{figm3}, we could find that with the inhomogeneity $\alpha$ increases, in strong coupling region, the ferromagnetic spin susceptibility $\chi(\Gamma)$ increases, and the antiferromagnetic spin susceptibility $\chi(K)$ decreases.

\begin{figure}[htpb]
\centerline {\includegraphics*[width=3.0in]{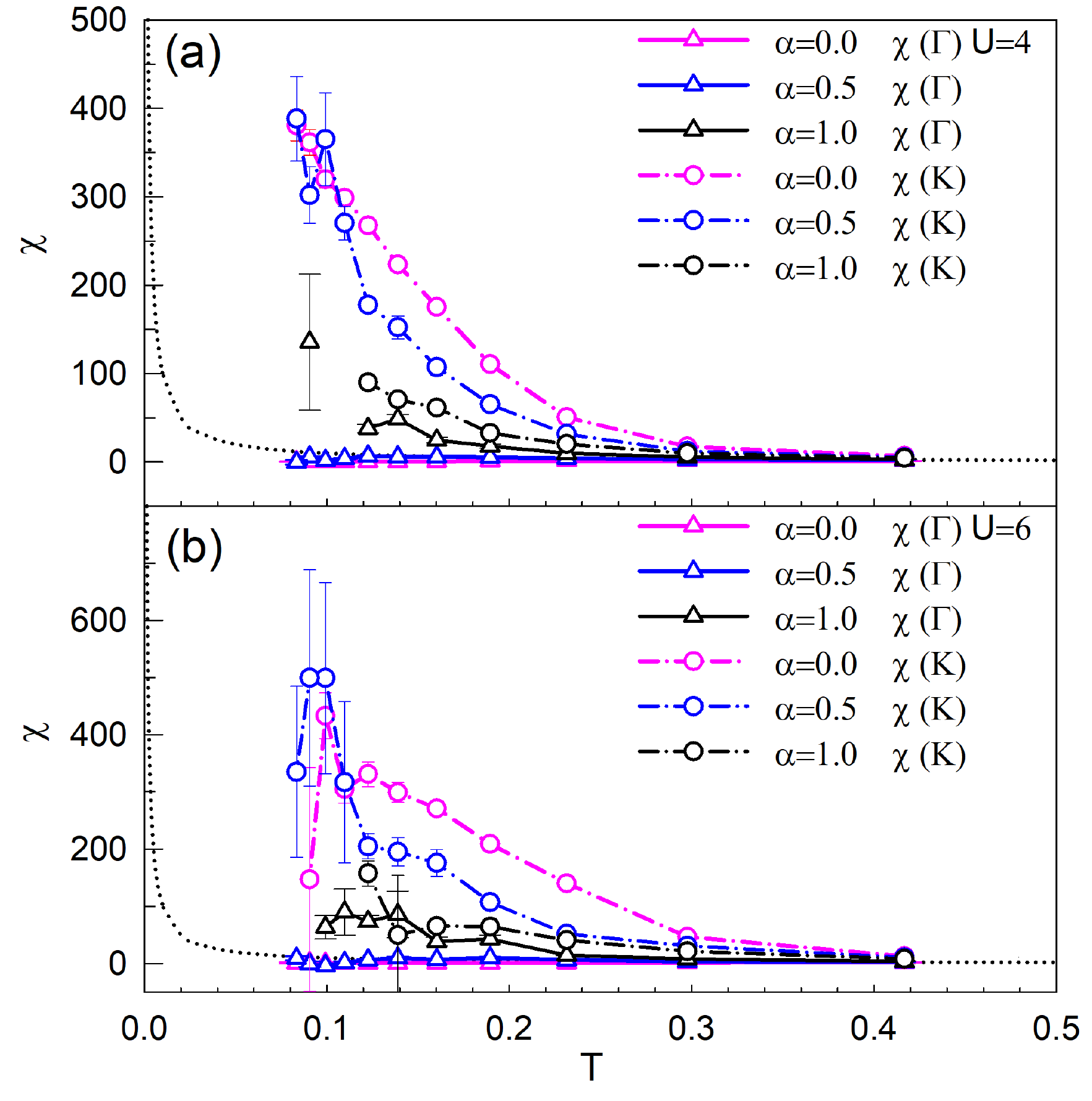}}
\caption{(Color online) Antiferromagnetic and ferromagnetic spin susceptibility $\chi(K)$, $\chi(\Gamma)$ are presented for different values of the inhomogeneity $\alpha$ and temperature $T$ at $L=8$. Results are obtained at (a)$U$=4 and (b)$U$=6.}
\label{figm4}
\end{figure}

Thus, there is a suppression of the spin susceptibility at $q=K$ point as $\alpha$ increases, and the spin susceptibility grows at $q=\Gamma$. However, the value of $\chi(k)$ is almost twice as large as $\chi(\Gamma)$  for the $U$ and temperature considered in Fig. \ref{figm3}. This means that although there is such suppression in $\chi(k)$, the nature of the system is still antiferromagnetic.
It is interesting to ask whether there should be a set of points $(U, T)$  where the ferromagnetic spin susceptibility will surpass the antiferromagnetic one. In Fig. \ref{figm4}, $\chi(K)$ and $\chi(\Gamma)$ are plotted vs $T$ for different values of $\alpha$ with $U=4, 6$.
One could find that the ferromagnetic spin susceptibility $\chi(\Gamma)$ is smaller than antiferromagnetic $\chi(K)$, so the antiferromagnetism is dominant in a large parameters region. From current results, one can see that, $\chi(K)$ is always larger than $\chi(\Gamma)$ except for case of $U=10$, $\alpha=1$ where $\chi(K)$ is almost as large as $\chi(\Gamma)$. Due to the limit of DQMC technique, the numerical instability prevent us to perform lower temperature or larger interaction, and it is difficult for us to conclude whether there would be a set of points $(U, T)$ where the ferromagnetic spin susceptibility will surpass the antiferromagnetic one. Anyway, the antiferromagnetism is dominant in a large parameters region.

\begin{figure}[htpb]
\centerline {\includegraphics*[width=3.0in]{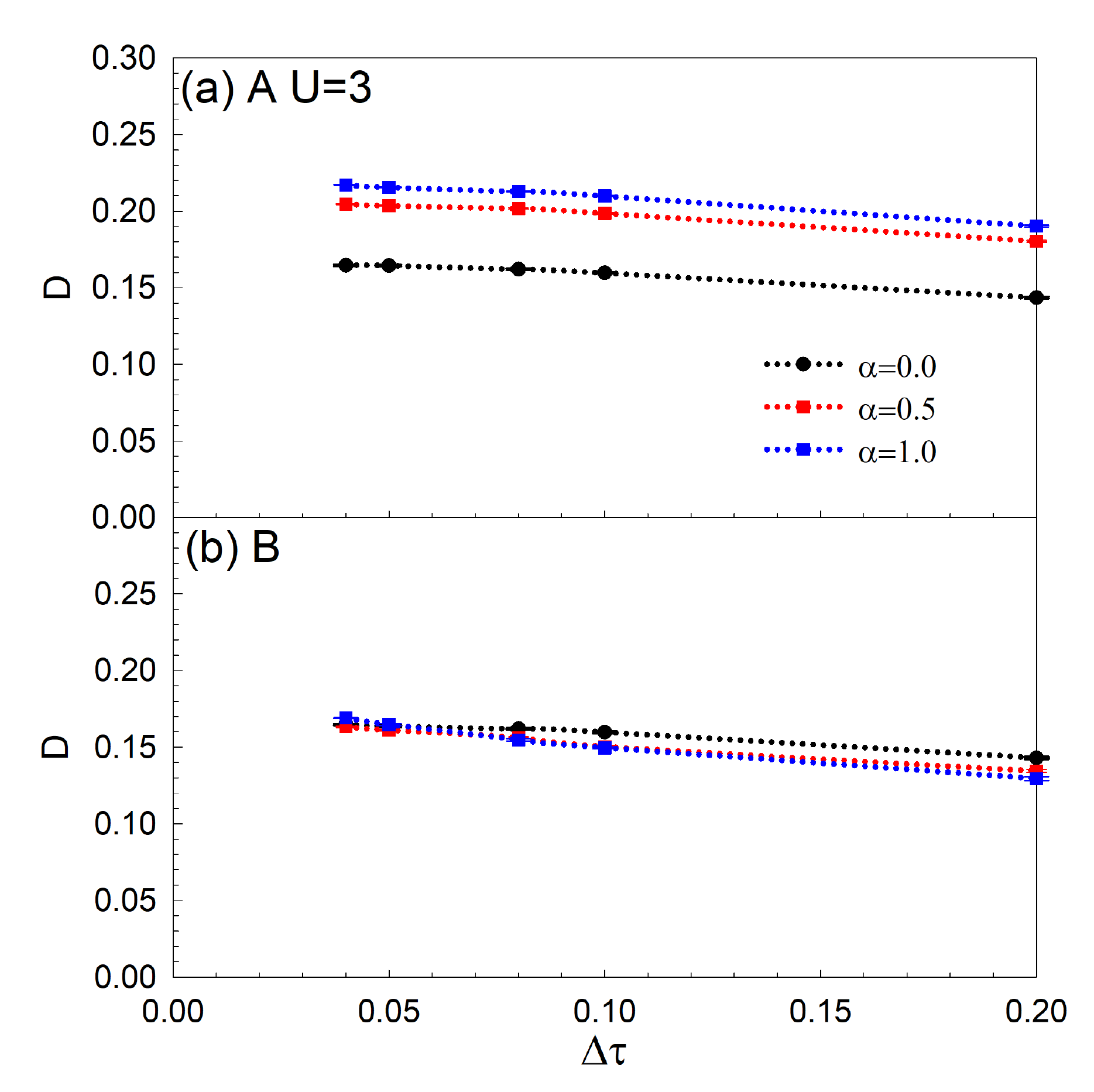}}
\caption{(Color online)(a) Double occupancy on $A$ sites for varying inverse temperature $\beta$ with different Trotter step $\Delta\tau$ at $U=3$, $\beta=6$ and $L=8$, (b) double occupancy on $B/C$ sites.}
\label{Figm13}
\end{figure}

\subsection{Trotter step}
\label{appendix:C}
In the DQMC algorithm, the systematic error mainly comes from the Trotter step, $\Delta\tau$. Fig. \ref{Figm13} shows double occupancy for different Trotter step $\Delta\tau$. One could find that double occupancy change slightly when Trotter step is smaller than 0.1. Due to the convergence of the finite $\Delta\tau$ scaling, we use the value of $\Delta\tau$ = 0.1 in all the simulations.

\bibliography{referencelieb}

\end{document}